# Nested Orthogonal Arrays

Joel Atkins[1] and David B. Zax

Department of Chemistry & Chemical Biology

Baker Laboratory

Cornell University

Ithaca, NY 14853

Abstract:

Orthogonal Arrays allow us to test various levels of each factor and balance the different factors so that we can estimate interactions as well as first order effects. There is a trade-off between how well we can sample different levels of each factor and how many interactions we are able to estimate. This paper describes one method to mitigate this trade-off. This method will allow us, with *n* observations, to sample *n* levels of each factor and minimize the correlation between the estimates of first order terms and their interactions.

1. Introduction

Statistics, arguably, came of age with Fisher's[7] development of experimental designs for agricultural experiments. As agriculture experiments are awkward candidates for extensive variations in methods of treatment (i.e. differing types and quantities of fertilizer and/or watering schedules for every observation) they provided an ideal testing ground for fractional factorial and balanced incomplete block designs[4].

A simple design with four points at (-1,-1), (-1,1), (1,-1), and (1,1) allows us to estimate two main effects and an interaction term as:

---

[1] Correspondence can be sent to Joel.Atkins@yahoo.com. Dr. Atkins now works at CNA Insurance.

$$\begin{bmatrix} B_{x_1} \\ B_{x_1} \\ B_{x_1 x_2} \end{bmatrix} = \begin{bmatrix} -0.25 & -0.25 & 0.25 & 0.25 \\ -0.25 & 0.25 & -0.25 & 0.25 \\ 0.25 & -0.25 & -0.25 & 0.25 \end{bmatrix} \begin{bmatrix} y_{-1,-1} \\ y_{-1,1} \\ y_{1,-1} \\ y_{1,1} \end{bmatrix}$$

The three vectors (-0.25, -0.25, 0.25, 0.25), (-0.25, 0.25, -0.25, 0.25), and (0.25, -0.25, -0.25, 0.25) are orthogonal and all of the elements have the same absolute value. This means that we have minimized the variance of their errors and that their errors are uncorrelated, if the errors are independent and identically distributed.

This design will not let us estimate any non-linear effects. For that, we will need to test additional levels for each variable. The four points (-1, 0.33), (-0.33, 1), (0.33, -1), and (1,-0.33) would let us estimate quadratic and even cubic effects for one of the variables. However, these points would not be ideal for estimating any interaction, since every point has $x_1 x_2 < 0$. (We would need more than four points to estimate higher order main effects and an interaction, this example is illustrative.)

In this paper, we will offer a novel method for designing nested orthogonal arrays. These designs will allow us estimate higher order main effects and interactions.

2. Orthogonal Arrays

Owen[15] was motivated by LHS designs to develop a scheme based on Orthogonal Arrays. While LHS designs seek to provide good coverage for each variable individually, Orthogonal Array (OA) designs seek to provide balance between any set of $t$ independent parameters, where $t$ is the strength of the OA design. In this scheme, the sample space is divided into $s^d$ hypercubes where $s^t/n$. An $OA(n,d,s,t)$ design is used to allocate the sample points to $n$ of the $s^d$ hypercubes, so that if the sample space is projected onto the space spanned by any $t$ variables, each of the $s^t$ hypercubes in this subspace will include $n/s^t$ sample points. Unfortunately, as the strength of the orthogonal array increases, $s$ will decrease (as we endeavor to keep $s^t$ within a small factor of $n$).

Thus, there is a trade-off between providing a good balance between larger sets of variables and the degree of precision for additive terms (and interactions between fewer variables). This trade-off can be appreciated when one realizes that an LHS design is an *OA(n,d,n,1)* design.

After constructing an orthogonal array, mapping the elements *{0, . . . , s-1}* onto independent random permutations of *{0, . . . , s-1}* for each variable, and choosing the locations of the points from uniform distributions over the resulting hypercubes gives unbiased estimates. Owen[15] showed that the average value of a function at the sample points approaches the average value of the function over the space at a rate of $n^{-(t+2)/2t}$ for terms involving at most *t* independent variables.

3. Nested Orthogonal Arrays

We can accomplish this by modifying an OA design. An OA design will place $s^{t-1}$ values of each variable into each of *s* strata. Placing each of these values into $s^{t-1}$ distinct smaller strata within their common larger strata will provide the structure of an LHS design while preserving the structure of an OA design. As an example, consider the design consisting of the four points 000, 011, 101, and 110, an *OA(4,3,2,2)* design. By mapping the 0's for each variable to a 0 and a 1 and mapping the 1's for each variable to a 2 and a 3, we get an *OA(4,3,4,1)* design. One such design would consist of the points 010, 132, 303, and 221. By considering the strata [0,1] and [2,3], this design can be viewed as an *OA(4,3,2,2)* design; and by considering 0, 1, 2, and 3 separately this design can be viewed as an *OA(4,3,4,1)* design. We will refer to such designs as Nested Orthogonal Arrays of strength 2, as this design has a strength 1 OA (or an LHS design) nested within a strength 2 OA. Tang[17] used this approach to construct designs which are both LHS designs and OA designs of strength 2.

A more ambitious goal is to construct a strength 3 NOA design, a design which contains

Orthogonal Arrays of strength 1, 2, and 3. The design in Table 1 is an *OA(64,5,8,2)* design. By considering the strata [0,1], [2,3], [4,5], and [6,7], it can be viewed as an *OA(64,5,4,3)* design. By mapping the 0's in each column onto the set [0, . . . ,7], the 1's in each column onto [8, . . . ,15], . . . , and the 7's in each column onto [56, . . . , 63], this becomes an *OA(64,5,64,1)* design. Thus, this mapping would give a strength 3 NOA.

Table 1

| | | | |
|---|---|---|---|
| 11111 | 21364 | 51427 | 61652 |
| 13333 | 23146 | 53605 | 63470 |
| 15555 | 25720 | 55063 | 65216 |
| 17777 | 27502 | 57241 | 67034 |
| 00357 | 30122 | 40661 | 70414 |
| 02175 | 32300 | 42443 | 72636 |
| 04713 | 34566 | 44225 | 74050 |
| 06531 | 36744 | 46007 | 76272 |
| 01573 | 31706 | 41045 | 71230 |
| 03751 | 33524 | 43267 | 73012 |
| 05137 | 35342 | 45401 | 75674 |
| 07315 | 37160 | 47623 | 77456 |
| 10735 | 20540 | 50203 | 60076 |
| 12517 | 22762 | 52021 | 62254 |
| 14371 | 24104 | 54647 | 64432 |
| 16153 | 26326 | 56465 | 66610 |

3.1 Constructing a Nested Orthogonal Array

Bush[1] provided a method for generating Orthogonal Arrays of the form *OA($s^t$,s+1,s,t)* when *s* is a prime power. This is done using polynomials in the finite Galois Field, *GF(s)*. The $s^t$ points in the sample are represented by the $s^t$ polynomials in *GF(s)* of order less than *t*. Elements of the first column, $x_{i1}$, are the coefficients of $y^{t-1}$ in the polynomial associated with the *ith* point. Elements of the other columns, $x_{ij}$, are determined by evaluating the polynomial associated with the *ith* point at the *(j-1)st* element of *GF(s)*. As an example, we consider an *OA($4^2$,5,4,2)* design. Each of the 16 points corresponds to one of the polynomials in *GF(4)* of order 0 or 1. We consider the row corresponding to *ax+(a+1)*. The element of the first column is the coefficient of *x*, or *a*. The elements of the other four columns are *ax+(a+1)* evaluated at *x* = *0, 1, a,* and *(a+1)*, respectively. These are *(a)(0)+(a+1) = (a+1), (a)(1)+(a+1) = 1, (a)(a)+(a+1) = 0,* and *(a)(a+1)+(a+1) = a*. We will use this method to help construct strength 3 NOA designs.

In our case, we want to create a design which can be viewed as an *OA($k_3s_3^3$,d,$s_3$,3)* design, an *OA($k_2s_2^2$,d,$s_2$,2)* design, and an *OA($s_1$,d,$s_1$,1)* design, with $s_1 \geq s_2 \geq s_3$. We will provide a method for constructing such a design with the maximum possible values of $s_1$, $s_2$, and $s_3$, when $p^4/n$ for some prime factor *p*. (When there is no prime *p*, such that $p^4/n$, we are left with either the trivial solution $s_2=s_3$ or the trivial solution $s_3=1$.)

Let $s_3$ to be the largest prime power such that $s_3^3/n$. Using Bush's method, it is possible to construct an *OA($s_3^3$,$s_3$+1,$s_3$,3)*. We discard the first column of this design, for reasons which will be discussed later (thus our scheme requires $d \leq s_3$). Appending $k_3$ repetitions of the remaining columns will form an *OA($k_3s_3^3$,d,$s_3$,3)* design. This design is already an *OA(($k_3s_3$)$s_3^2$,d,$s_3$,2)* design, but it will be possible to incorporate an *OA($k_2s_2^2$,d,$s_2$,2)* design with $s_2>s_3$ into our *NOA($k_3s_3^3$,d,3)* design if there exists some $c \geq 1$ such that $p^{2c}/k_3s_3$ and $p^c+1 \geq d$. This will give better resolution for any two term interactions.

Choose *b* and *c* to maximize *c* while satisfying $bp^{2c}=k_3s_3$. It is then possible to form an *OA($p^{2c}$,d,$p^c$,2)* design using Bush's method (hence the requirement that $p^c+1 \geq d$), and append *b* copies of this design to form an *OA($k_3s_3$,d,$p^c$,2)* design. Each of the rows of this design will be associated with $s_3^2$ contiguous rows of the *OA($k_3s_3^3$,d,$s_3$,3)* design. Within each set of $s_3^2$ rows, each row will belong to the same repetition of the original *OA($s_3^3$,d,$s_3$,3)* design and have the same element in the first column of the original *OA($k_3s_3^3$,$s_3$+1,$s_3$,3)* design. At this point, we will be able to form a new design by multiplying the elements of the *OA($k_3s_3^3$,d,$s_3$,3)* design by

$p^c$ and adding the corresponding elements from the $OA(k_3s_3,d,p^c,2)$ design (with these operations done in the field of integers).

Clearly, this design is an $OA(k_3s_3^3,d,s_3,3)$ design, when viewed in terms of the appropriate strata. It is necessary to show that is also an $OA(k_2s_2^2,d,s_2,2)$ design, where $s_2=p^cs_3$. We can do this by showing that each pair of elements from the $OA(k_3s_3,d,p^c,2)$ design is added to each pair of elements from the $OA(k_3s_3^3,d,s_3,3)$ design $b$ times. In light of how the designs were combined, it is sufficient to show that each pair of elements from the $OA(k_3s_3^3,d,s_3,3)$ design is included once in each of the $k_3s_3$ sets of $s_3^2$ rows used to assign the rows from the $OA(k_3s_3,d,p^c,2)$ design. This can be seen by recalling that the rows in each set are from one repetition of a $OA(k_3s_3^3,s_3+1,s_3,3)$ design and all have the same element of the first column of the original $OA(k_3 s_3^3,s_3+1,s_3,3)$ design. Sharing a common element in the first column means that the remaining columns of these $s_3^2$ rows will form an $OA(s_3^2,s_3,s_3,2)$ design. Thus, every pair of columns must contain each pair of elements exactly once in each of these sets of rows. This is the reason for our earlier exclusion of the first column of Bush's $OA(k_3s_3^3,s_3+1,s_3,3)$ design.

Now, it is only necessary to incorporate an $OA(n,d,n,1)$ design. The new design contains $bp^cs_3$ repetitions of each of the elements in the set $[0, \ldots, p^cs_3-1]$. Now, for each element $i$ within each column, we assign each of the $bp^cs_3$ repetitions of the value $i$ distinct values in the range $[bp^cs_3 i, \ldots, bp^cs_3(i+1)-1]$. This is done in a random manner to avoid bias. This mapping will form a $OA(n,d,n,1)$ design and, as the larger strata were preserved, we now have a $NOA(n,d,3)$ design.

In creating the design shown in Table 1, the original $OA(64,5,4,3)$ design was constructed using Bush's method. Bush's method was also used to create an $OA(4,3,2,2)$ design which made it possible to expand the last three columns of the $OA(64,5,4,3)$ design and form an $OA(64,3,8,2)$ design. Based on the *OA(4,3,2,2)* design an *OA(64,5,2,2)* design was constructed by visual inspection. This design made it possible to include the first two columns of the *OA(64,5,4,3)* design. Visual inspection may be a possible means of including $s_3+1$ independent variables in small experiments.

To avoid bias, the elements (0, 1, ..., $s_i$-1) of each column in each of the $k_3$ repetitions of the *OA($s_3^3$,d,$s_3$,3)* design and each of the *b* repetitions of the *OA($p^{2c}$,d,$p^c$,2)* design are mapped onto random permutations of (0, . . . , $s_i$-1) (with all of the permutations chosen independently.) To minimize variance, the rows of the *OA($k_3s_3$,d,$p^c$,2)* design are shuffled randomly before the two designs are combined.

4. Conclusions

Combining the coverage of individual variables which LHS offers with the balance provided by Orthogonal Arrays can give powerful designs. Nested Orthogonal Arrays also combine the strengths of Orthogonal Arrays and the strengths of Latin Hypercube Sampling. In the simulations we have done, the NOA designs have consistently outperformed strength 2 OA designs and LHS designs. We expect that this will always be the case, due to the multiple nested arrays.

Acknowledgements

The preliminary research in this area was performed with the support of the ACS-Petroleum Research Fund under grant #31204-AC6.

References

1 Bush, K., (1952) Orthogonal Arrays of Index Unity. *Annals of Mathematical Statistics*, **23** 426-434.

2 Cheng, V., Suzukawa, H., and Wolfsberg, M. (1973). Investigations of a Nonrandom

Numerical Method for Multidimensional Integration. *Journal of Chemical Physics*, **59** 3992-3999.

4 Davies, O., et. al. (1956). *The Design and Analysis of Industrial Experiments.* Oliver & Boyd, London.

55 Davis, P. and Rabinowitz, P. (1984). *Methods of Numerical Integration*, 2nd Edition. Academic Press Inc, Orlando.

6 Fang, K. and Wang, Y. (1984). *Number-theoretic Methods in Statistics*, Chapman and Hall, New York.

7 Fisher, R. (1925). *Statistical Methods for Research Workers*. Oliver & Boyd, Edinburgh.

8 Haselgrove, C. (1961). A Method For Numerical Integration. *Mathematics of Computation*, **15** 323-337.

9 Hlawka, E. (1962). Zur Angenäherten Berachnung Mehrfacher Integrale. *Monatsh. Math*. **66**, 140-151.

10 Hlawka, E. Funktionen von Beschränkter Variation in der Theorie der Gleichverteilung. *Ann. Mat. Pura Appl*. **4-54**, 325-333.

11 Hua, L. and Wang, Y. (1981). *Applications of Number Theory to Numerical Analysis*, Springer-Verlag and Science Press, Berlin and Beijing.

12 McKay, M., Conover, W., and Beckman, R. (1979). A Comparison of Three Methods for Selecting Values of Input Variables in the Analysis of Output from a Computer Code. *Technometrics* **21**, 239-245.

13 Niederreiter, H. (1992). *Random Number Generation and Quasi-Monte Carlo Methods*. SIAM, Philadelphia.

14 Owen, A. (1992). A Central Limit Theorem for Latin Hypercube Sampling. *Journal of the*

*Royal Statistical Society* **54**, 541-551.

15 Owen, A. (1992). Orthogonal Arrays for Computer Experiments, Integration and Visualization. *Statistical Sinica* **2**, 439-452.

16 Owen, A. (1997). Scrambled Net Variance for Integrals of Smooth Functions. *Annals of Statistics* **25**, 1541-1562.

17 Tang, B. (1993). Orthogonal Array-Based Latin Hypercubes. *Journal of the American Statistical Association* **88**, 1392-1397

18 Zaremba, S. (1968). The Mathematical Basis for Monte Carlo and Quasi-Monte Carlo Methods. *SIAM Review* **10**, 303-314.